 \documentclass[article]{svmult}
\usepackage{amssymb}
\usepackage{amsmath}
\usepackage{mathrsfs}
\usepackage[sort,numbers]{natbib}

\usepackage{epsfig}
\usepackage{graphics}
\usepackage{float}
\usepackage{tabularx}
\usepackage{color}
\newcolumntype{C}[1]{>{\centering\arraybackslash}p{#1}}
\setcitestyle{square}
\begin{document}
 

\title*{Galactic Center gamma-ray excess and Higgs-portal Dark Matter}
\author{Tanmoy Mondal and Tanushree Basak}
\institute{Tanmoy Mondal \at Theoretical Physics Division, Physical Research Laboratory, Ahmedabad 380009, India \& 
Department of Physics, Indian Institute of Technology, Gandhinagar, Ahmedabad, India, \email{tanmoym@prl.res.in} \and
Tanushree Basak \at Theoretical Physics Division, Physical Research Laboratory, Ahmedabad 380009, India, \email{tanu@prl.res.in}}
\maketitle

\def\be{\begin{equation}}
\def\ee{\end{equation}}
\def\al{\alpha}
\def\bea{\begin{eqnarray}}
\def\eea{\end{eqnarray}}
\def\beas{\begin{eqnarray*}}
\def\eeas{\end{eqnarray*}}

\abstract{
 From astronomical observations, we know that dark matter exists and makes up $\sim$25\% of our Universe. 
Recently the study of anomalous gamma-ray emission in the regions surrounding the galactic center has 
drawn a lot of attention. It has been pointed out that the excess of 1-3 GeV gamma-ray in the low 
latitude is consistent with the emission expected from annihilating dark matter. I will discuss the 
Higgs-portal dark matter models which can explain these phenomena because of the presence of scalar 
resonance. In addition, the parameter space of these models also satisfy constraints from the LHC Higgs 
searches, relic abundance and direct detection experiments. The gauged $U(1)_{B-L}$ model is very well suited with 
the  FERMI-LAT observation along with other constraints.}

\title*{Galactic Center gamma-ray excess and Higgs-portal Dark Matter}
\author{Tanmoy Mondal and Tanushree Basak}
\institute{Tanmoy Mondal \at Theoretical Physics Division, Physical Research Laboratory, Ahmedabad 380009, India \& 
Department of Physics, Indian Institute of Technology, Gandhinagar, Ahmedabad, India, \email{tanmoym@prl.res.in} \and
Tanushree Basak \at Theoretical Physics Division, Physical Research Laboratory, Ahmedabad 380009, India, \email{tanu@prl.res.in}}

\maketitle

\section{Introduction}

Center of our galaxy possesses large dark matter (DM) density and is a very good target for indirect detection. 
Recently the observation of anomalous Gamma-ray from the galactic center (GC) and the inner galaxy regions by  Fermi-LAT 
 has gained a lot of attention. 
The data shows a spatially extended excess of $\sim 1-3$ GeV gamma rays from the regions surrounding the galactic center. 
Annihilations of a few $\times10$ GeV WIMP (weakly interacting massive particle) 
DM candidate annihilating to quarks or leptons ($\tau^+\tau^-$) can fit the data. Apart from DM interpretation several attempts were 
made to explain the excess as astrophysical origin with unresolved millisecond pulsars near the GC. 
But the observed excess extends upto $\sim 10^\circ$ which disfavors the 
astrophysical origin of the excess. 
More recently an analysis of the spectrum of the anomalous gamma-ray has confirmed that the signal is very 
well fitted by a 31-40 GeV DM particle annihilating to $b\bar b$ with 
an annihilation cross section of $\sigma v = (1.4-2.0)\times 10^{-26} \textrm{cm}^3 \textrm{sec}^{-1}$ 
\cite{Daylan:2014rsa} and references therein.

  In this article, we have shown that Higgs-portal dark matter model can explain the observed gamma-ray excess.
  We have considered the minimal $U(1)_{B-L}$ extension of the SM. 
  The third generation right handed neutrino being odd under $\mathbb{Z}_2$-symmetry serves as
  a viable DM candidate~\cite{Basak:2013cga}. 
  This model is consistent with the requirements to explain the GC $\gamma$-ray 
  excess along with constraints coming from LHC bound on SM-Higgs, relic density and direct 
  detection of DM. 

 \section{Minimal $U(1)_{B-L}$ gauge extension of SM}
\label{sec:b-l-dm}
The model under consideration is the minimal gauged  $U(1)_{B-L}$ extension of the SM, for details see 
\cite{tanmoy} and references therein. This model contains a singlet scalar $S$ with $B-L$ charge +2, three right-handed 
neutrinos $N_R^i (i=1,2,3)$ with $B-L$ charge -1 along with SM fields. $\mathbb{Z}_2$ charge of the third generation RH neutrino,
$N_R^3$ is odd and hence qualified as a viable DM candidate~\cite{Basak:2013cga}.

 Scalar potential of this model reads as:
\be\label{BL-potential}
V(\Phi,S) = 
m^2\Phi^{\dagger}\Phi + \mu ^2|S|^2 + \lambda _1 (\Phi^{\dagger}\Phi)^2 +\lambda _2|S|^4 + \lambda _3 \Phi^{\dagger}\Phi|S| ^2 
\nonumber \,.
\ee 
$\Phi$ and $S$ are the doublet and singlet scalar fields, respectively. After symmetry breaking the
scalar fields can be written as, $S=\frac{v_{_{B-L}}+\phi'}{\sqrt{2}}$ and $\Phi=(0\;\;\frac{v+\phi}{\sqrt{2}})^T$.
 The two scalar mass eigenstates can be written in terms of gauge eigenstates with mixing angle $\alpha$,
\begin{eqnarray}
 H_2 &=& \sin\alpha \; \phi' + \cos\alpha \; \phi \; , \\
 H_1 &=& \sin\alpha \; \phi - \cos\alpha \; \phi' ,
\end{eqnarray}
where, $H_2$ is identified as the SM-Higgs boson and we consider $m_{H_2}>m_{H_1}$.

Interaction of the DM and the Higgses originate from the following term, 
\begin{eqnarray}
\label{yuk}
 \mathcal{L}_{int} &=& \sum_{i=1}^3\frac{y_{n_i}}{2}\overline{N_R^i} S N_R^i\;.
\end{eqnarray}
Mass of dark matter is given by, $m_{_{DM}}=m_{N_R^3}=\frac{y_{n_3}}{\sqrt 2}v_{_{B-L}}$.

\subsection{Constraints from LHC}
Discovery of Higgs boson at LHC will constrain scalar mixing angle severely. The signal strength of a particular channel reads as: 
\be \label{eq:ratio}
r_i^{xx}  = \frac{\sigma_{H_i}}{\sigma_{H_i} ^{SM}}\cdot \frac{BR_{H_i \to xx}}{BR_{H_i \to xx} ^{SM}}\;, \; (i=1,2).
\ee 
where, $\sigma_{H_i}$ and $BR_{H_i \to xx}$ are the production cross section of $H_i$ , and the branching ratio of
$H_i \to xx$ respectively. For the SM Higgs the corresponding quantities are $\sigma_{H_i} ^{SM}$ and $BR_{H_i \to xx} ^{SM}$.
The invisible decay width of the SM Higgs reads as
\be\label{eq:H2inv}
\Gamma_{H_2}^{Hid}\equiv \Gamma_{inv} = \frac{m_{H_2} \,\lambda_{DM}^2}{16\pi} \sin^2\alpha \left(1-4\frac{m_{_{DM}}^2}{m_{H_2}^2} \right)^{\frac{3}{2}},
\ee
Since $\lambda_{DM} (\equiv y_{n_3})$ is suppressed by large $B-L$ breaking VEV, $\Gamma_{inv}$ remains very small ($\sim 0.5\%$). 

In order to realize $H_2$ as a SM Higgs, we need $r_2 \geq 0.9$ (0.8)
and correspondingly $r_1 \leq 0.1$ (0.2).  From Fig.~\ref{fig:b-l-relic-SI}(right panel) we found that $r_2$ 
being $\geq 0.9$ (0.8) restricts the choice of scalar mixing such that $\cos\alpha \geq 0.96$ (0.94) for $m_{_{DM}}\sim 31$ GeV.



 \subsection{Breit-Wigner enhancement}

In general the annihilation of Majorana fermionic DM into SM-fermion pairs through a scalar mediator is velocity 
suppressed. In that case the thermally averaged annihilation cross-section can be written as,
\begin{equation*}
 \langle \sigma v\rangle = a +bv^2 \; , \textrm{where $a,b$ are model dependent variables.}
\end{equation*}
The term $a$ comes from s-channel s-wave process, where as, $b$ has contributions from
both s-wave and p-wave. The averaged velocity $v$ can be expressed as, $v\sim \sqrt{3/x}$. Because of p-wave suppression, 
$\langle \sigma v\rangle$ at the time of freeze-out ($x_f\sim 20$) is different than that at the 
galactic halo ($x\sim 10^6$). However, $\langle \sigma v\rangle$ at the galactic halo can be substantially enhanced using 
the Breit-Wigner mechanism \cite{murayama,guo}, where the DM annihilates through a narrow s-channel resonance. 

The leading annihilation channels of DM are, 
${N_{R}^3} {N_{R}^3} \longrightarrow b\bar{b},\,\tau^+ \tau^-$.
The s-channel resonant annihilation cross-section into final state $b\bar{b}$ (dominant) is given as,
\be
 4E_1E_2\;\sigma v = \frac{1}{8\pi}\sqrt{1-\frac{4m_b^2}{s}}|\bar{\mathcal{M}}|^2 \nonumber \\
                 = \frac{\lambda_{DM}^2\cos^2\alpha}{32\pi^2}\frac{s^2}{m_{H_1}^2}\frac{m_{H_1}\Gamma_{H_1}}{(s-m_{H_1}^2)^2+m_{H_1}^2\Gamma_{H_1}^2},
\ee
where, $\Gamma_{H_1}$ is the total decay width of $H_1$.

Here, we introduce two parameters $\delta$ and $\gamma$ as,
\begin{equation}
m_{H_1}^2=4m_{_{DM}}^2(1-\delta)\; , \; \gamma=\Gamma_{H_1}/m_{H_1}.
\end{equation}
Adopting the single-integral formula, we obtain,
\be
\label{eq:bw}
 \langle \sigma v\rangle = \frac{1}{n_{EQ}^2}\frac{m_{_{DM}}}{64\pi^4x}\int_{4m_{_{DM}}^2}^\infty ds\;(4E_1E_2\sigma v g_i^2 ) \sqrt{s}
  \times\;\sqrt{1-\frac{4m_{_{DM}}^2}{s}}\;K_1\!\left(\frac{x\sqrt{s}}{m_{_{DM}}}\right).
\ee
where, $n_{EQ}=\frac{g_i}{2\pi^2}\frac{m_{_{DM}}^3}{x}K_2(x)$.
$K_1(x)$ and $K_2(x)$ are the modified Bessel's function of second kind and $g_i$ is the internal degrees of freedom of dark matter
particle.

\begin{figure}[t!]
 \centering
 \includegraphics[width=5.4cm, angle =0]{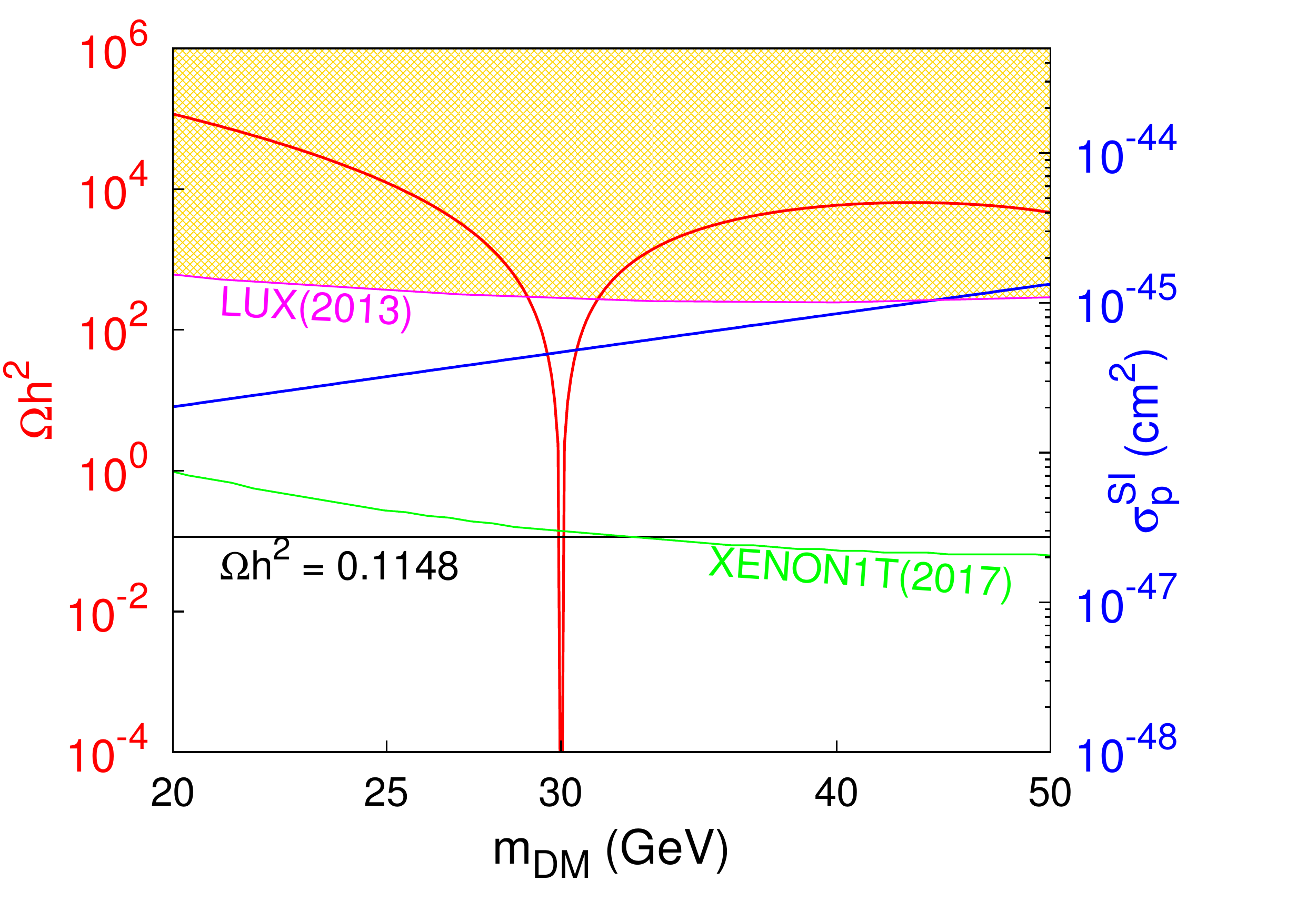}
 \includegraphics[width=3.9cm, angle =0]{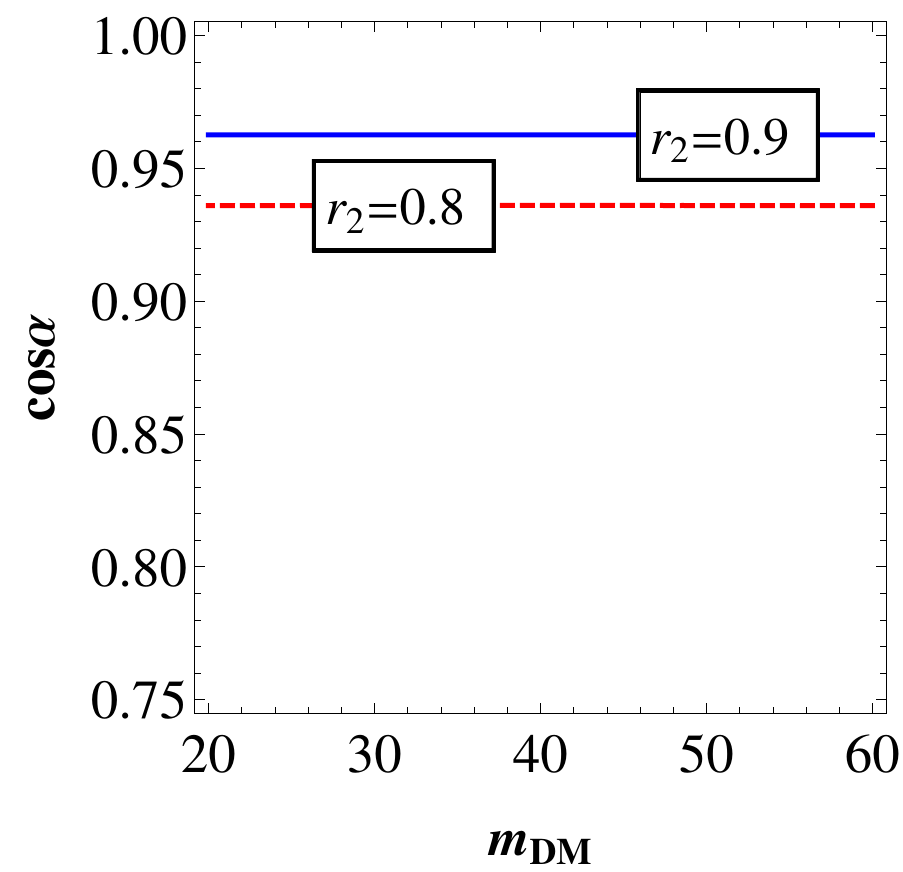}
\caption{(Left panel)Relic abundance (red curve) and scattering cross-section (blue curve) as a function of DM mass. 
LUX(2013) excluded region is shown in yellow  and the green line shows future sensitivity of {\sc Xenon1T} experiment.
(Right panel)Contours of signal strength in $\cos\alpha-m_{_{DM}}$ plane. }
 \label{fig:b-l-relic-SI}
\end{figure}

We again redefine $s$ as, $s=4m_{_{DM}}^2(1+y)$ where, $y\propto v^2$. 
Eq.\ref{eq:bw} can be recast in terms of $\delta$, $\gamma$ and $y$ as,
\begin{equation}
 \langle \sigma v\rangle\propto x^{3/2}\int_0^{y_{eff}}\frac{\sqrt{y}(1+y)^{3/2}e^{-xy}}{(y+\delta)^2+\gamma^2(1-\delta^2)}dy\;,
\end{equation}
where, $y_{eff}\sim \textrm{max}[4/x, 2|\delta|]$ for $\delta<0$ and $y_{eff}\sim 4/x$ for $\delta>0$ case. 
If $\delta$ and $\gamma$ are much smaller than unity, $\langle \sigma v\rangle$ scales as $v^{-4}$ in the
limit $v^2\gg\textrm{max}[\gamma,\delta] $. 
%
%
%
Fig.\ref{fig:b-l-relic-SI} shows the relic abundance (red curve) as a function of 
DM mass. The resultant relic abundance is found to be consistent with the reported value of WMAP-9 \cite{wmap9} 
only near resonance when, $m_{_{DM}} \sim (1/2)\; m_{H_1}$.

We have also achieved the required $\langle \sigma v\rangle_{b\bar{b}} \sim 1.881\times10^-26\; \textrm{cm}^3/s$ at the galactic halo 
through the Breit-Wigner enhancement given the value of parameters $\delta\simeq -10^{-3}$ and $\gamma \simeq 10^{-5}$. Note that, 
the same set of parameter values have been used to compute the relic abundance. 

\subsection{Constraints from direct detection searches}
The scattering cross-section (spin-independent) for the dark matter off a proton or neutron is given as,
\begin{equation}
\label{eq:sigmaSI}
 \sigma_{p,n}^{SI}=\frac{4m_r^2}{\pi}f_{p,n}^2\;,
\end{equation}
where, $m_r$ is the reduced mass defined as, $1/m_r=1/m_{_{DM}}+1/m_{p,n}$ and  $f_{p,n}$ is the hadronic matrix element. 
The f-values are given in \cite{Ellis:2000ds}. Here, $a_q$ is the effective coupling constant between the DM and the quark. 
An approximate form of $a_q/m_q$ can be recast as :
\begin{eqnarray}
\label{eq:aqmq}
 \frac{a_q}{m_q} &=& \frac{\lambda_{DM}}{v\sqrt{2}} \Bigg[\frac{1}{m_{H_1}^2}-\frac{1}{m_{H_2}^2}\Bigg]
 \sin\!\alpha \cos\!\alpha \; .
\end{eqnarray}
In Fig.\ref{fig:b-l-relic-SI}  the yellow region above is excluded by LUX(2013) \cite{lux}. 
We found that the spin-independent scattering cross-section (blue curve) value satisfies LUX limit and {\sc Xenon1T} 
experiment \cite{xenon1T} (green line) may constrain the model in near future.


\section{Summary}
\label{sec:summary}

The excess of $\gamma$-ray emission in the low latitude region near the 
galactic center can be explained by annihilation of DM (in the mass range $\sim 31-40$ GeV) into $b\bar b$, with cross-section 
of the order of the weak cross-section. 
Here we have analyzed a Higgs-portal DM model, namely $U(1)_{B-L}$ model and constrain the parameter space of this model. 
RH-neutrino DM in this case is well-suited for explaining the galactic-center gamma-ray excess 
along with satisfying other DM and collider constraints. Because of a narrow scalar resonance the required $\langle \sigma v \rangle$ was obtained through Breit-Wigner enhancement mechanism. Future experiment like {\sc Xenon1T} can restrict 
the parameter space of minimal $U(1)_{B-L}$ model.  

%
%

\end{document}